\documentclass[referee]{raa}            

\usepackage{graphicx,times}             
\input{epsf.sty}                        

\begin{document}

 \title{On-board GRB trigger algorithms of SVOM-GRM
\footnotetext{$*$ Supported by 973 Program 2009CB824800, NSFC10978001 and the Knowledge Innovation Program
of the Chinese Academy of Sciences, under Grant No. 200931111192010.}
}

   \volnopage{Vol.0 (200x) No.0, 000--000}      
   \setcounter{page}{1}          

   \author{Dong-Hua Zhao
      \inst{1}
      \inst{2}
   \and Bo-Bing Wu
      \inst{1}
   \and Li-Ming Song
      \inst{1}
   \and Yong-Wei Dong
      \inst{1}
   \and St\'{e}phane Schanne
      \inst{3}
   \and Bertrand Cordier
      \inst{3}
   \and Jiang-Tao Liu
      \inst{1}      
   }
   
\institute{$^1$ Institute of High Energy Physics, Chinese Academy of Sciences, Beijing 100049, China; {\it zhaodh@ihep.ac.cn}\\
$^2$ National Astronomical Observatories, Chinese Academy of Sciences, Beijing 100012, China;\\
$^3$ CEA Saclay, DSM/Irfu/Service d'Astrophysique, 91191, Gif-sur-Yvette, France }
\date{Received~~2009 month day; accepted~~2009~~month day}

\abstract{
GRM (Gamma-Ray Monitor) is the high energy detector on-board the future Chinese-French satellite SVOM (Space-based multi-band astronomical Variable Object Monitor)
which is dedicated to Gamma-Ray Burst (GRB) studies. This paper presents the investigation of the on-board counting rate trigger algorithms of GRM.
The trigger threshold and trigger efficiency based on the given GRB sample are calculated with the algorithms.
The trigger characteristics of GRM and ECLAIRs are also analyzed. In addition, the impact of solar flares on GRM is estimated, and the method to distinguish solar flares from GRBs is investigated.
}

   \authorrunning{D.-H. Zhao, B.-B Wu, L.-M. Song, Y.-W. Dong, S. Schanne \& B. Cordier}            
   \titlerunning{On-board GRB trigger algorithms of SVOM-GRM}  

   \maketitle

%
%
\section{Introduction}           
\label{sect:intro}

Gamma-ray bursts (GRBs), which are flashes of gamma-rays associated with extremely energetic explosions in the cosmological distance,
are unpredictable and short. And the satellite telemetry band-passes are not large enough to send every detection event to the ground in real time.
In addition, the satellite will encounter many non-GRB events in orbit which can introduce false triggers in detector.
These factors make it important to develop on-board trigger algorithms for three purposes: 1) to detect GRBs as early as possible which makes
it possible to detect the early afterglow and to allow the follow-up observations of the GRBs with high redshift; 2) to increase the trigger efficiency
of GRBs and the types of GRBs; 3) to reject the false triggers rate and to decrease the false trigger.

A uniform and easily understood trigger algorithm, which searched for count increases over a few time scales and the backgrounds were estimated
by taking an average of the count rate from a period before the burst, was employed by most previous experiments (vela, PVO, ISEE-3, Ginga and BATSE).
In such a trigger algorithm, a large threshold ($\geq11\sigma$) was usually set to decrease the false triggers. However, even with such a large threshold,
most experiments still had a high false trigger rate.
The statistical fluctuations and the trends of the background are the two main factors causing false triggers, and the latter one becomes the critical factor
when using large threshold. In addition to the traditional trigger algorithm, the High Energy Transient Explorer (HETE-2) (Fenimore et al.~\cite{Fenimore2001}; Tavenner et al.~\cite{Tavenner2001})
estimates the background in the trigger sample region by fitting the background regions with polynomial. Such trigger algorithms can remove the trends effectively
and allows HETE-2 to use a much smaller threshold than previous experiments. The Burst Alert Telescope (BAT) of SWIFT
(Fenimore et al.~\cite{FenimoreD2001}; Fenimore et al.~\cite{Fenimore2004}; Mclean et al.~\cite{Mclean2004}) also adopts the similar trigger algorithms.

SVOM is a LEO mission with an altitude of 600 km and an inclination of 30$^\circ$.
Like SWIFT, SVOM (Space-based multi-band astronomical Variable Object Monitor) (Paul et al.~\cite{Paul2011}; Basa et al.~\cite{Basa2008}) can slew rapidly the low energy instruments (MXT and VT)
to the sources for follow-up observations. Before starting this operation, the satellite needs to confirm that there is a burst localized by the high energy
payloads GRM (Gamma-Ray Monitor) and ECLAIRs.
Both GRM and ECLAIRs can get trigger information by the counting rate trigger algorithm, and ECLAIRs can further confirm the source and
obtain its location by the image trigger algorithm (Schanne et al.~\cite{Schanne2008}).

GRM is sensitive in the energy range from ~30 keV to ~5 MeV. It is a phoswich detector which consists of three scintillator layers of plastic scintillator, NaI(Tl) and CsI(Na).
The three kinds of scintillators with a diameter of 190 mm are glued together and viewed with the same light guide coupled to a PMT.
The thicknesses of plastic scintillator, NaI(Tl) and CsI(Na) are respectively 6 mm, 15 mm and 35 mm.
NaI(Tl) works as the main detecting element of GRM while CsI(Na) is another important detection element and also serves
as anti-coincidence element against photons from behind. The plastic scintillator is dedicated to reject the background events due to the low energy charged particles.
The beryllium plate with a thickness of 1.5 mm is chosen as entrance window of this triple phoswich detector.
And the collimator made of tantalum is located in front of the scintillator case to reduce the background by limiting the FOV to 2.5 steradians.
ECLAIRs is a coded-mask imaging camera for X- and gamma-rays with a 2 sr-wide FOV and 1024 cm$^2$ detector area. It mainly consists of a detection plane of 80$\times$80 CdTe semiconductor detectors,
a coded-mask located above the detection plane, a multi-layer lateral shield between the detection plane and the mask, and some mechanical structures, etc.
Its detection energy range is 4 keV - 250 keV. We build the mass model of GRM and ECLAIRs with Geant4 package.
And for the detailed descriptions of the instruments, see the references (Zhao et al.~\cite{Zhao2012}; Dong et al.~\cite{Dong2009};
Godet et al.~\cite{Godet2009}; and Mandrou et al.~\cite{Mandrou2008}), respectively.

In this study, we investigate the on-board GRB trigger algorithms of GRM in detail based on the given GRB sample and estimate the impact of solar flares on GRM.
The paper is organized as follows: The GRB sample used in the trigger algorithms is briefly described in Section 2. We study three counting rate trigger algorithms,
try to find the most sensitive energy ranges and time scales, and compare the trigger characteristics of GRM and ECLAIRs in Section 3.
In order to seek for the method to reject false triggers, we analyze the impact of solar flares on GRM and investigate
how to distinguish the triggers caused by solar flares from those by GRBs in Section 4. Finally, in Section 5, we conclude with a concise summary and
a discussion of some limitations in this work as well as some further studies required.

\section{The GRB sample}

We collect a GRB sample consisting of 249 long GRBs and 103 short GRBs for the investigation of the trigger algorithms.
All of the spectra we used are the time-integrated spectra.

The spectra of long GRBs are from Table 9 in (Kaneko et al.~\cite{Kaneko2006}). And these GRBs are selected from the 2704 GRBs
of BATSE based on the criterion of a peak photon flux in 256 ms (50-300 keV) greater than 10 photons/s/$\rm cm^2$ or a total energy fluence in the summed energy range
($\sim$20-2000 keV) larger than $2.0\times10^{-5}$ ergs/$\rm cm^2$. A set of photon models were used to fit each GRB in the energy range 30-2000 keV,
and the best-fit model was chosen according to $\chi^2$ probabilities and parameter constraints. Excluding the GRBs which have no duration information\footnote{http://gammaray.msfc.nasa.gov/batse/grb/catalog/current/tables/duration\_table.txt} or no flux/fluence data, we get 249 long GRBs.

\begin{figure}[!ht]
\centering
\includegraphics[height=0.25\textheight, width=0.7\textwidth]{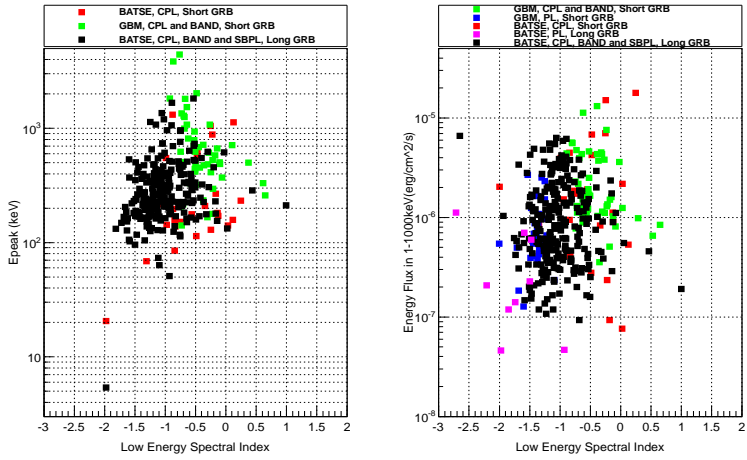}	
\caption{The distribution of the GRB sample on Low energy spectral index vs. $E_{\rm peak}$ and on Low energy spectral index vs. Energy flux.}
\label{Fig:lowindex-Epeak}
\end{figure}

For the 103 short GRBs, 26 GRBs are from BATSE (Ghirlanda et al.~\cite{Ghirlanda2004}) and the other 77 short GRBs are from GBM (Nava et al.~\cite{Nava2011}).
In (Ghirlanda et al.~\cite{Ghirlanda2004}), the authors selected the short GRBs ($T_{90}\leq$2 sec) with a peak flux (computed on 64 ms
timescale and integrated over energy range 50-300 keV) exceeding 10 photons/cm$^2$/s from the GRB catalog\footnote{http://cossc.gsfc.nasa.gov/cossc/batse/}
and fitted them with the comptonization model (Ghirlanda et al.~\cite{Ghirlanda2002}) in the energy range $\sim$30 keV-1.8 MeV. A sample of 28 short GRBs with 100-300 keV
fluence $\geq2.4\times10^{-7}$ erg/$\rm cm^2$ was obtained. Excluding the two GRBs with uncertain $E_{\rm peak}$, we finally get the 26 short GRBs.
In (Nava et al.~\cite{Nava2011}), the spectra of GRBs detected by GBM up to March 2010 were analyzed with single power-law model, Band function
and Comptonized model which are defined in (Kaneko et al.~\cite{Kaneko2006}). And 77 short GRBs with the best-fit model were obtained.

The parameter distributions of the GRB sample are shown in Fig.~\ref{Fig:lowindex-Epeak}.
The spectra of short GRBs with the average low energy spectral index of -0.72 are harder than that of long GRBs with index of -1.07.
The average $E_{\rm peak}$ of short GRBs (exclude the GRBs with spectra of PL) is 662 keV which is larger than that of long GRBs (294 keV).
 And the average energy flux ($2.3\times10^{-6}$ erg/$\rm cm^2$/s) of short GRB is approximately two times of that
of long GRBs ($1.2\times10^{-6}$ erg/$\rm cm^2$/s). 

The long and short GRBs described above are bright GRBs with raletively high peak photon flux and energy fluence. In addition, we get another set of GRBs, which are called dark GRBs thereafter, by reducing the flux of the bright GRBs and keeping other features the same.
Thus, we get a GRB sample including bright and dark GRBs to study the trigger algorithms of GRM. All of the light curves required are from the dataset of
BATSE\footnote{http://www.hep.csdb.cn/browsall/compton/data/batse/ascii\_data/64ms/}.
For the GRBs of GBM, we select the corresponding light curve of BATSE with the similar duration.

We get the detected spectra of GRBs by inputting the corresponding spectra into the mass model built with Geant4.
And then we can get the time information for each detected photon by random sampling according to the corresponding light curves.
Finally, we can obtain the detected photon lists with energy and time information which will be used as the input data to investigate the trigger algorithms of GRM and ECALIRs.

\section{The GRB triggers}

In this section, we study the GRB triggers of GRM with the counting rate trigger algorithms.
We describe three kinds of counting rate trigger algorithms, and compute the corresponding trigger threshold and trigger efficiency.
We also analyze the GRB triggers of ECLAIRs in the same way for comparisons.
NaI(Tl) is the main detection crystal of GRM. And the threshold and efficiency of GRM discussed in this section are the results of GRM\_NaI.

\subsection{Counting rate trigger algorithms}

The counting rate trigger algorithm is a method to look for GRBs by searching for a "significant" increase in the photon count rate over a background count rate.
And the photon count rate is generally corresponding to the given energy range, time scale and the detector plane zone. The detector plane zone is applicable to the
pixel detector but not to the scintillation detector like GRM. For the preliminary studies of the GRB trigger of ECLAIRs, we do not consider the detector plane zone.
Triggering in different energy ranges can tailor the detector sensitivity to hard or soft GRBs. And triggering on different time scales can adapt the detector to long or short GRBs.
We select 30-50 keV, 50-150 keV, 150-300 keV, 300-550 keV and 550-5000 keV as the trigger energy ranges for GRM and select 4-50 keV, 4-80 keV, 4-120 keV and 15-50 keV
for ECLAIRs considering the detection efficiency and the spectral feature of GRBs. One of the important reasons why the energy range is divided finely is that
we will try to reject false triggers by analyzing the distribution of triggers on energy ranges which is discussed in Section 4.
In addition, we need to prepare for the investigation of the cooperations between GRM and ECLAIRs on GRB triggers.
According to the experience from the previous satellites, we adopt 5 ms$\times2^n$
as the trigger time scales of GRM and ECLAIRs. We define 5 ms-80 ms as short time scales and 160 ms-40 s as long time scales.

Generally, in the trigger algorithm, the range of times when there is no apparent emission from GRB is the background period (thereafter Back) and that
when the GRB will probably produce strong emissions is the foreground period (thereafter Fore) (Fenimore et al.~\cite{FenimoreD2001}).
The algorithms on short and long time scales are called short rate trigger algorithm and long rate trigger algorithm, respectively.
In short rate trigger algorithm, the background during Fore is expressed by the average count rate in the Back in a period before
Fore (Fenimore \& Galassi~\cite{Fenimore2001}; Tavenner et al.~\cite{Tavenner2001}).
Such a algorithm is very appropriate for short GRB triggers in real time because its computation is simple and fast.
In the long rate trigger algorithm on long time scales, the period needed to compute the background is much longer
during which the trend of the background is obvious. In order to deal with the trend, generally people will fit
the background with a proper function. And the simple and effective linear function is used in our studies.
Then we can subtract effectively the background during Fore by interpolating or extrapolating the fitting function.
The long rate trigger in which all of the Backs are before the Fore is called long one-sided trigger, and that in which some Backs
are after the Fore is called long bracketed trigger. We denote the Backs before and after the Fore by Back1 and Back2, respectively.
The long bracketed trigger can subtract the background in foreground period best.
But it needs the longest time because the trigger has to be delayed until after Back2. In order to reduce the delay time,
Back2 is usually much shorter than Back1 and the space between Back2 and Fore is also shorter.
The parameter setting of the trigger algorithm is discussed in (Tavenner et al.~\cite{Tavenner2001}). In order to detect all kinds of GRBs in real time and
measure the spectra as complete as possible, we can run these three kinds of trigger algorithms on the on-board computers simultaneously.

\subsection{Trigger Threshold}

With the three kinds of trigger algorithms described above, we investigate the GRB triggers of GRM and ECLAIRs.
In (Zhao et al.~\cite{Zhao2012}), we simulated the background of GRM and ECLAIRs for seven typical Earth positions relative to the FOV.
With the same method and the given relative Earth positions on the orbit, we can get the background varying with the satellite orbit, namely with time.
In this study, we only consider one factor for the orbit, namely the Earth position.
For ECLAIRs, we take account three kinds of gamma-ray background. And for GRM, we also consider the delayed background caused by the trapped high energy protons in SAA as a constant approximately,
because it will not change obviously with the relative Earth position. From the background simulations, we know that ECLAIRs has more background counts than GRM in their own energy ranges,
and the background of ECLAIRs varies with the Earth position more obviously than that of GRM.

\begin{figure}[!ht]
\centering
\includegraphics[width=0.7\textwidth]{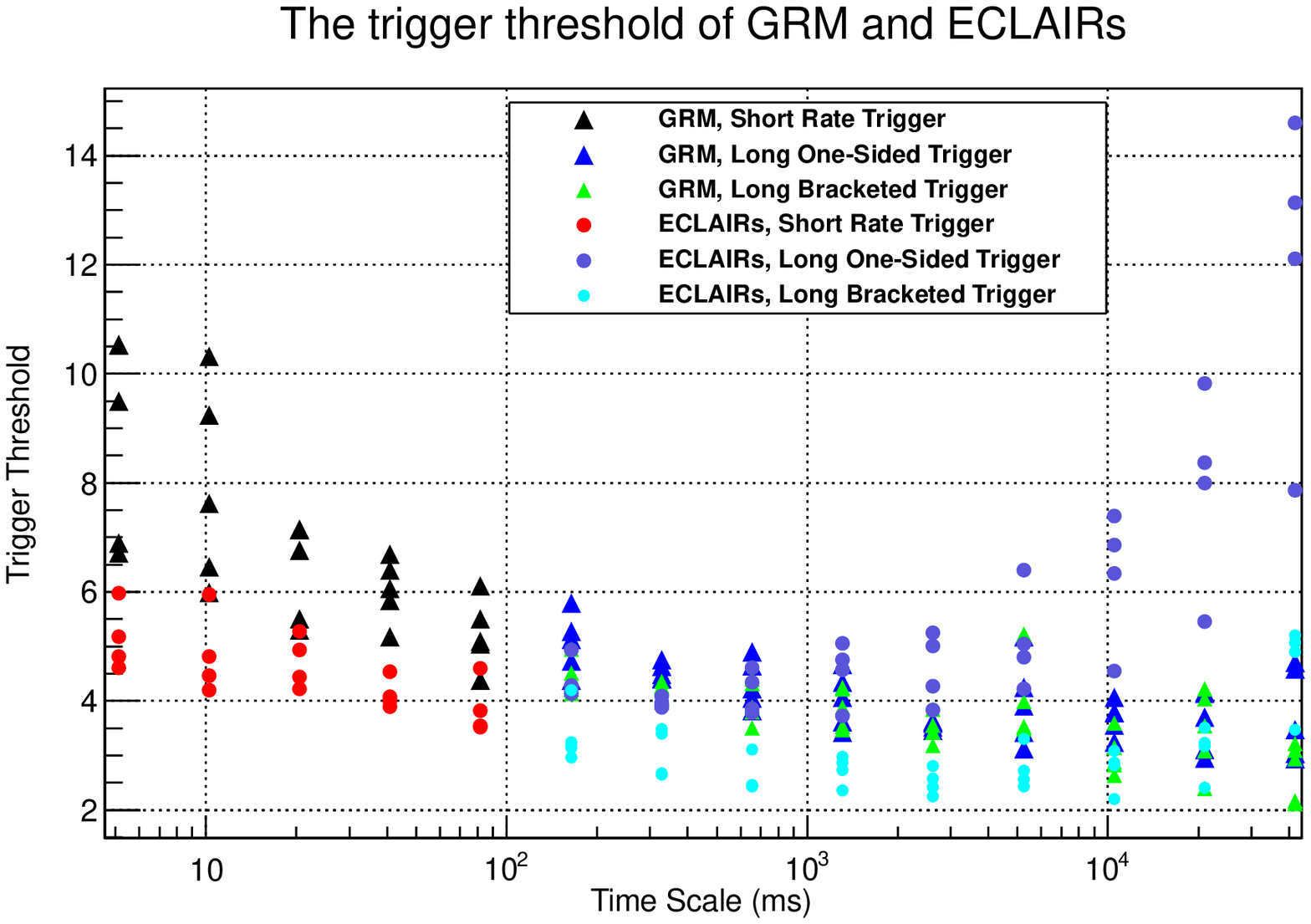}	
\includegraphics[width=0.7\textwidth]{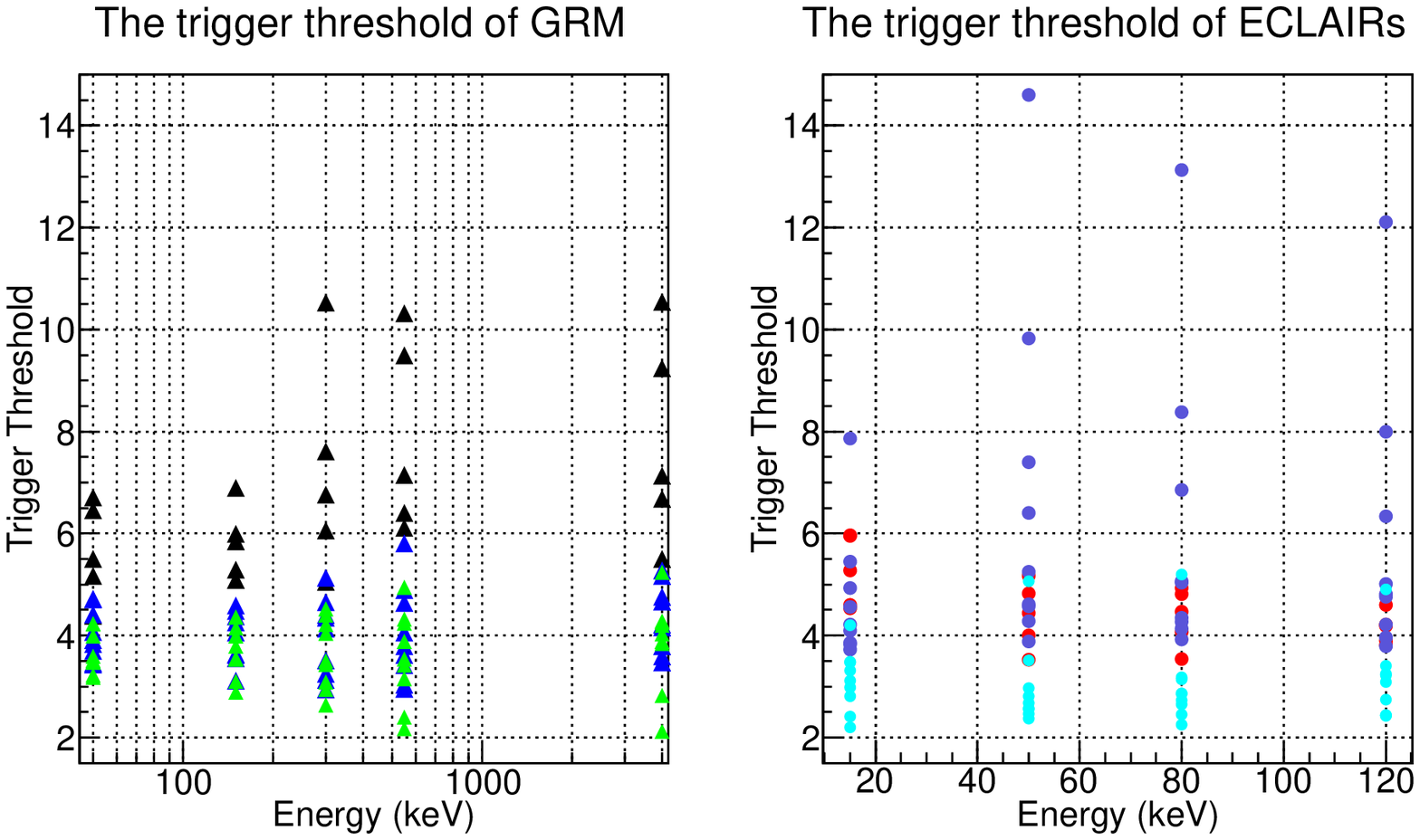}	
\caption{The trigger thresholds ($S_{max}$+0.5) of GRM and ECLAIRs on the different time scales (top) and in the different energy ranges (bottom).
For GRM and ECLAIRs, we use the up edge to express the corresponding trigger energy range (one exception: 15 keV in right bottom panel expresses the energy range 15-50 keV).}
\label{Fig:threshold}
\end{figure}

With the background, we can set the threshold which should cause neither too many false triggers nor too low trigger efficiency.
There are 70 (56) trigger energy range and time scale combinations for GRM (ECLAIRs).
And for each combination, we compute the threshold in the way inspired by the method provided in (Mclean et al.~\cite{Mclean2004}).
Firstly, we get the background varying with the satellite orbit corresponding to different combinations.
Then, we run the trigger algorithms with no injected bursts under the background conditions
to compute the corresponding maximum value $S_{max}$=(Fore-Back)/$\sqrt{\rm Back}$ for each combination.
Finally, we set the appropriate threshold according to $S_{max}$.
In order to avoid the false triggers due to the known background, we set $S_{max}$+0.5 as the threshold in this study.
The thresholds of GRM and ECLAIRs with different trigger algorithms in various energy ranges and on different time scales
are shown in Fig.~\ref{Fig:threshold}. The data of GRM and ECLAIRs are denoted by solid triangles and solid dots, respectively.
And the data due to different trigger algorithms are indicated by different colors.
In the top panel, the different values for each time scale are corresponding to different energy ranges.
In the bottom two panels, the different values for each energy range are corresponding to different time scales.
In order to analyze the threshold changing with the energy ranges simply and clearly,
we use the up edge to express the corresponding trigger energy range (one exception: 15 keV for ECLAIRs expresses the energy range 15-50 keV).
The same method of displaying data are also used in Fig. 3-Fig. 6. And the meanings of the markers in the figures are explained in the corresponding legends.

The threshold depends on the count rate, the trend of the background, as well as the trigger algorithm.
And comparing with the background of GRM, ECLAIRs has larger count rate and more rapid trend on their own time scale and energy range combinations.
The top panel in Fig. 2 shows that the threshold of GRM is much larger than that of ECLAIRs on the short time scale where the maximum threshold of GRM appears
(e.g. On the time scale 5 ms, the maximum threshold of GRM is 10.5 and that of ECLAIRs is 5.9.); that the threshold of ECLAIRs is very large with the long one-sided trigger
algorithm on the long time scale where its maximum threshold appears; and that the long bracketed trigger decreases the threshold of ECLAIRs obviously
(e.g. In the combination of 40 sec and 4-80 keV, the threshold of ECLAIRs with long one-sided trigger algorithm is 13 and that with long bracketed trigger is 5.).
As a result, we can conclude that: i) on the short time scale, the threshold is mainly determined by the count rate of the background and its fluctuation; and the threshold decreases as the count rate increases. ii) on the long time scale, the threshold mainly depends on the trend and the algorithm; and the threshold will be obviously smaller using long bracketed trigger than using 
long one-sided trigger if the trend is rapid. iii) on the long time scale, it is of great importance for ECLAIRs to adopt the long bracketed trigger algorithm to reduce the threshold to detecte dark GRBs. Comparing the threshold on time scales and that in energy ranges in Fig. ~\ref{Fig:threshold}, the largest threshold values of GRM lie in the energy ranges above 150 keV,
that is because the count rates in these energy ranges on short time scale are the lowest. And the largest threshold values of ECLAIRs lie in the energy ranges of 4-50 keV, 4-80 keV and 4-120 keV because of the most rapid trend of background and the use of long one-sided trigger.

\subsection{Trigger Efficiency}

With the trigger threshold, we can compute the corresponding trigger efficiency for the given GRBs described in Section 2 for each time scale and energy range combination.
In this section, we will use the dark GRBs which are obtained by reducing the flux of the bright GRBs to 1/10 or 1/100 and keeping other features the same.
And with the application of dark GRBs, the sensitive time scales and the sensitive energy ranges, as well as the characteristics of GRBs which are triggered in GRM and ECLAIRs
can be found out.	

\begin{figure}[!ht]
\centering
\includegraphics[width=0.7\textwidth]{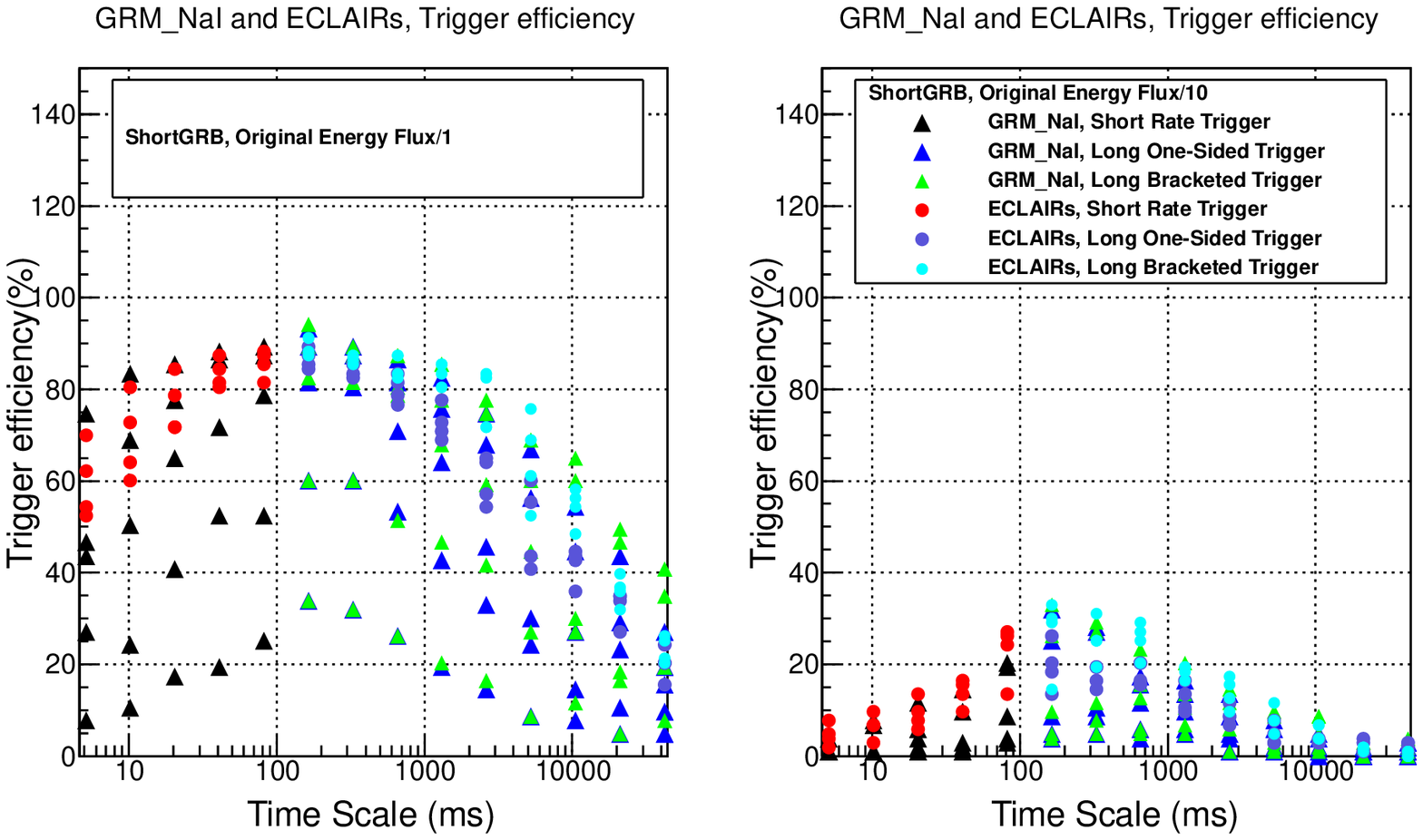}	
\includegraphics[width=0.7\textwidth]{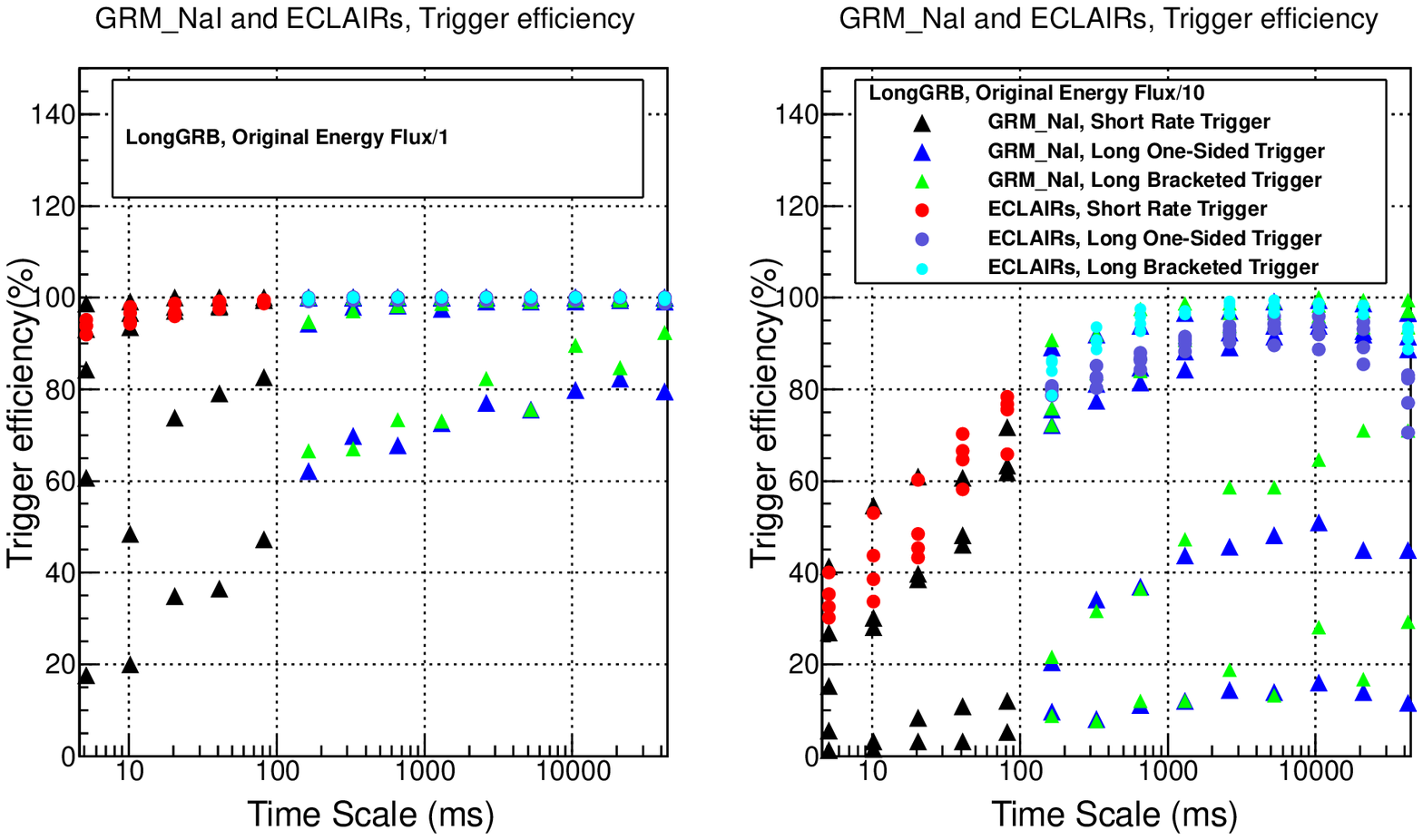}	
\caption{Top: The efficiencies of GRM and ECLAIRs for bright (left) and dark (right) short GRBs on different time scales.
Bottom: The efficiencies of GRM and ECLAIRs for bright (left) and dark (right) long GRBs on different time scales. }
\label{Fig:eff-time}
\end{figure}

The variations of the trigger efficiency for short and long GRBs with time scales are shown in Fig.~\ref{Fig:eff-time}.
We present the trigger efficiencies of GRM and ECLAIRs for bright GRBs in the left two panels and for dark GRBs in the right two panels.
The trigger efficiency for dark GRBs is obviously lower than that for bright GRBs overall.
Besides that, Fig.~\ref{Fig:eff-time} shows that:
1) On different time scales, the maximum efficiencies of GRM and ECLAIRs are similar (e.g. On time scale 1 sec, both the maximum efficiencies of
GRM and ECLAIRs are 85.4\% for bright short GRBs, and are 100\% for bright long GRBs);
2) On long time scales, the efficiency with long one-sided trigger is lower than that with long bracketed trigger which is more obvious for ECLAIRs
(e.g. On time scale 1 sec, the maximum efficiency is 91.5\% with long one-sided trigger for the long GRBs with 1/10 of the flux of bright long GRBs,
and is 96.7\% with long bracketed trigger) because of the higher trigger threshold using long one-sided trigger;
3) For long GRBs, GRM and ECLAIRs have higher efficiency on long time scales because the accumulation of the count rate on long time scales can increase their significance, and the efficiency on short time scales below 1 sec decreases quickly as the flux of GRBs decrease;
4) For short GRBs, GRM and ECLAIRs have the maximum efficiencies on a few hundred ms (The maximum efficiencies of ECLAIRs and GRM are  89.3\% and 93.2\%, respectively,
and both of them are on the time scale 160 ms.).

\begin{figure}[!ht]
\centering
\includegraphics[width=0.7\textwidth]{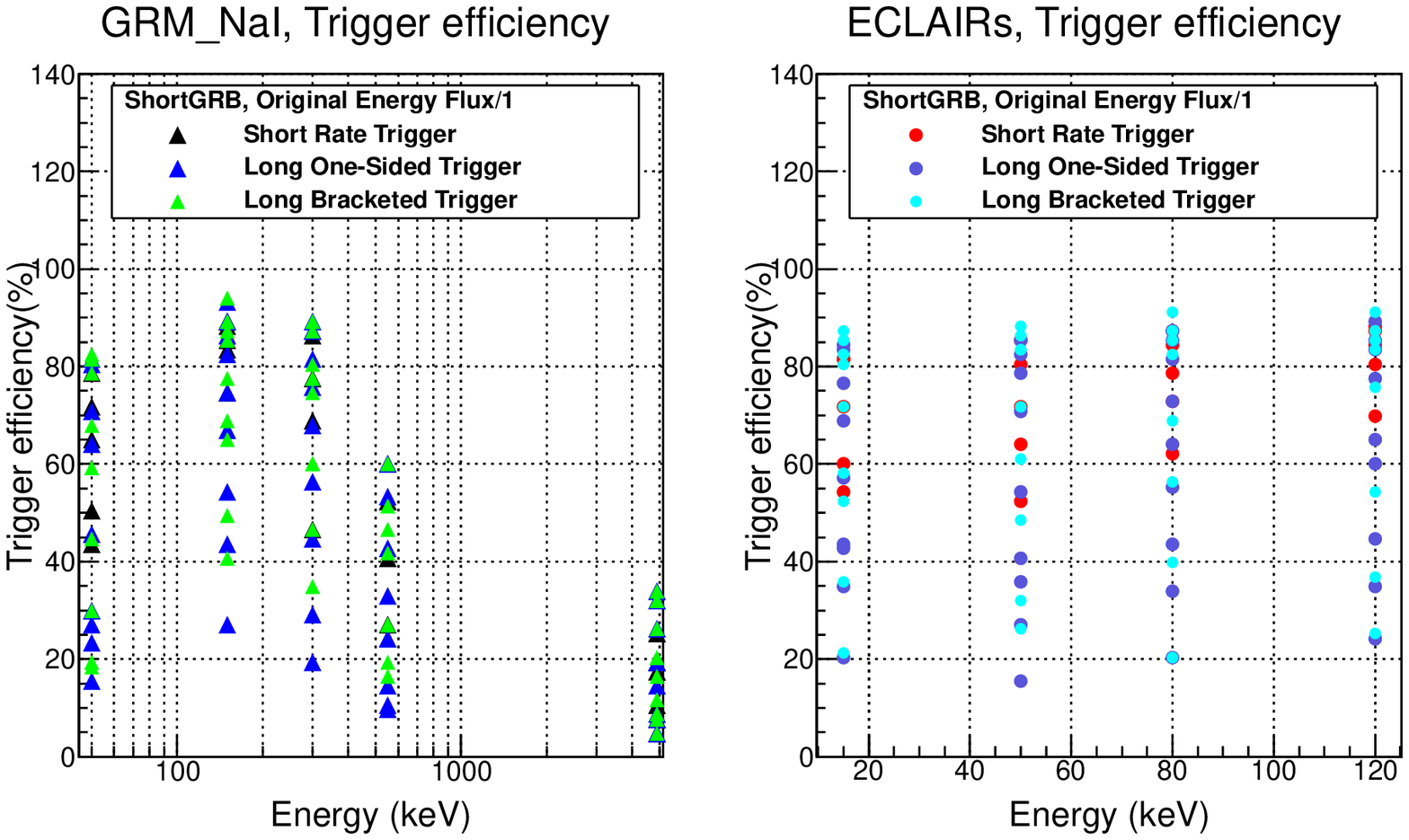}	
\includegraphics[width=0.7\textwidth]{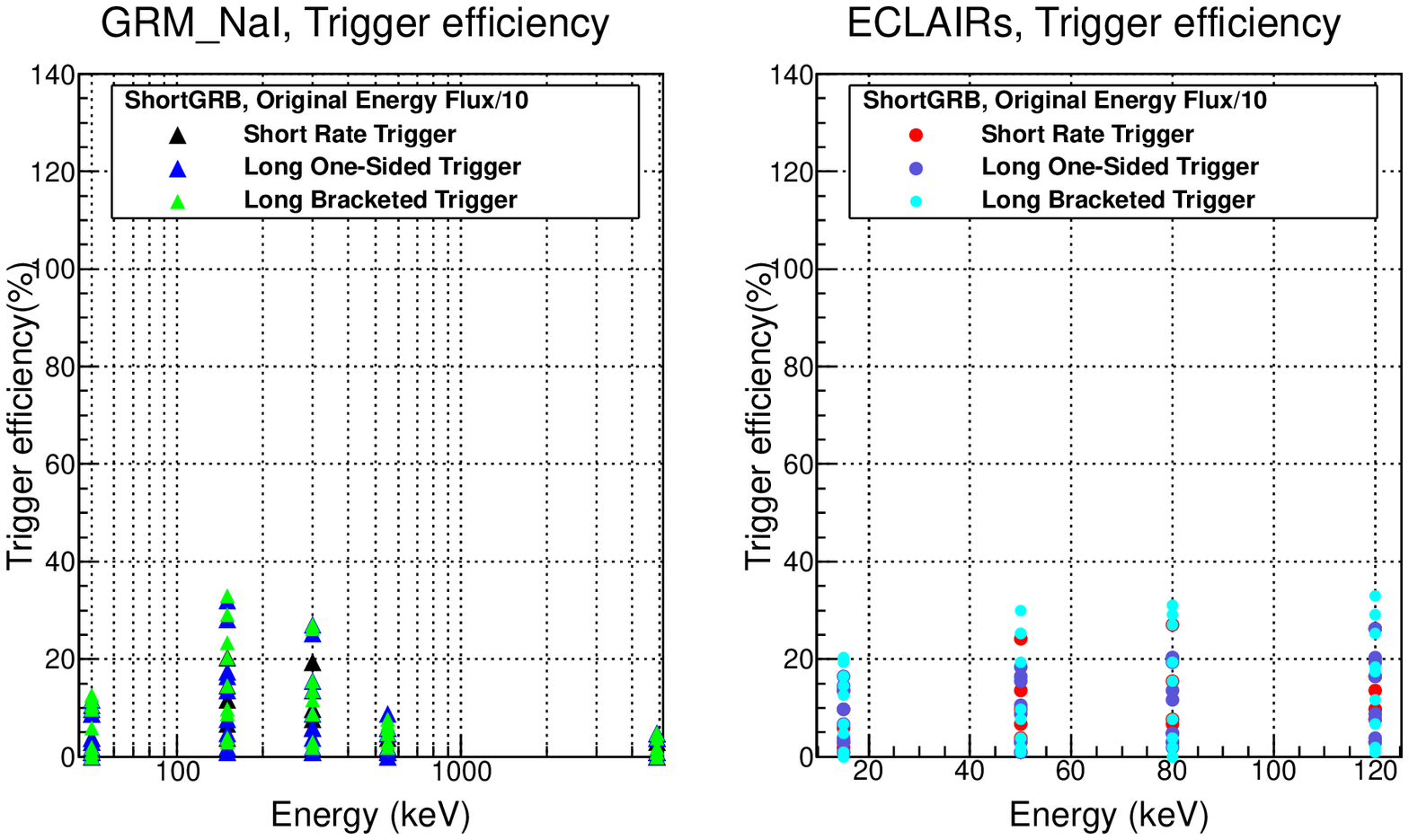}	
\caption{The efficiencies of GRM and ECLAIRs for bright (top) and dark (bottom) short GRBs in different energy ranges.}
\label{Fig:eff-energy-short}
\end{figure}

\begin{figure}[!ht]
\centering
\includegraphics[width=0.7\textwidth]{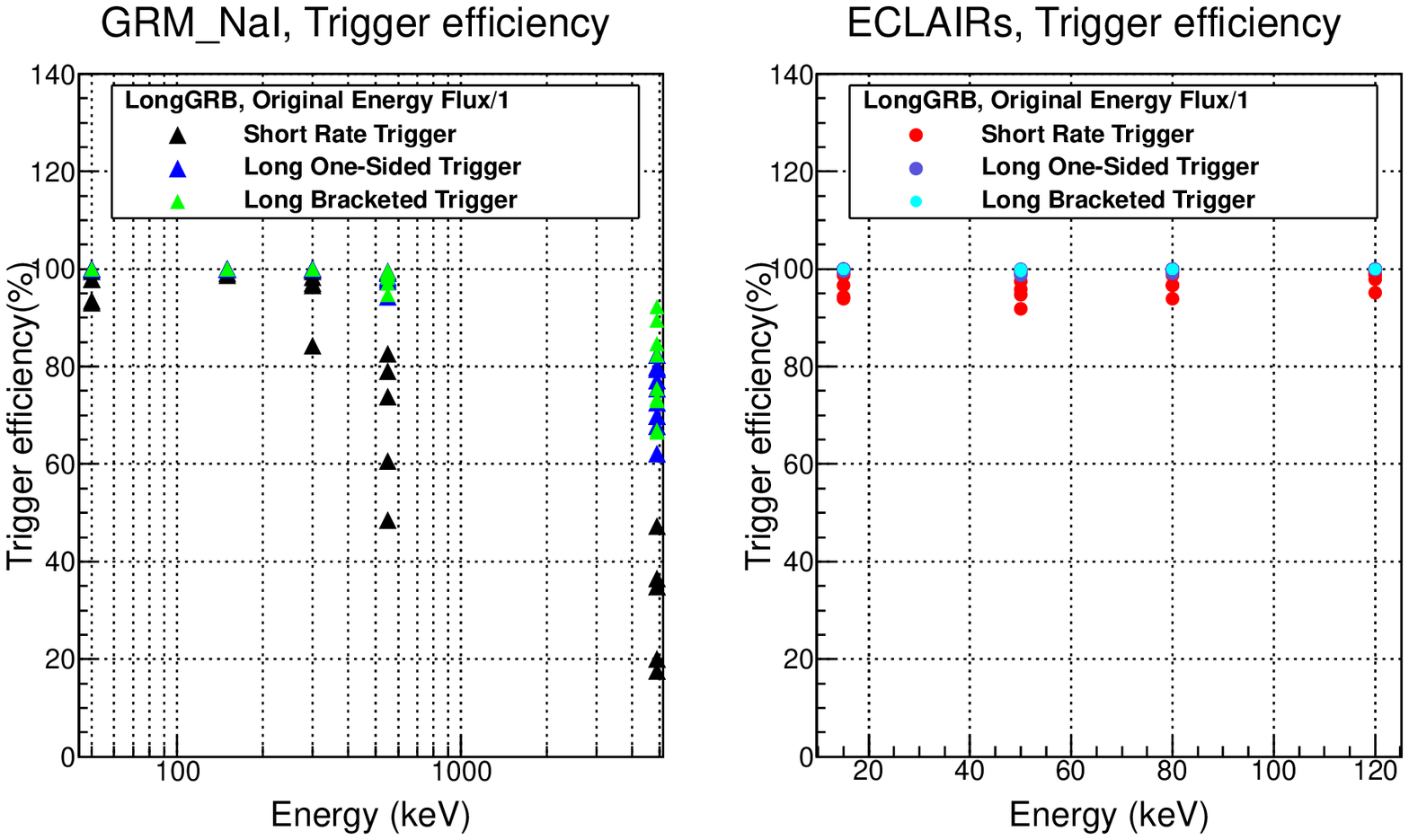}	
\includegraphics[width=0.7\textwidth]{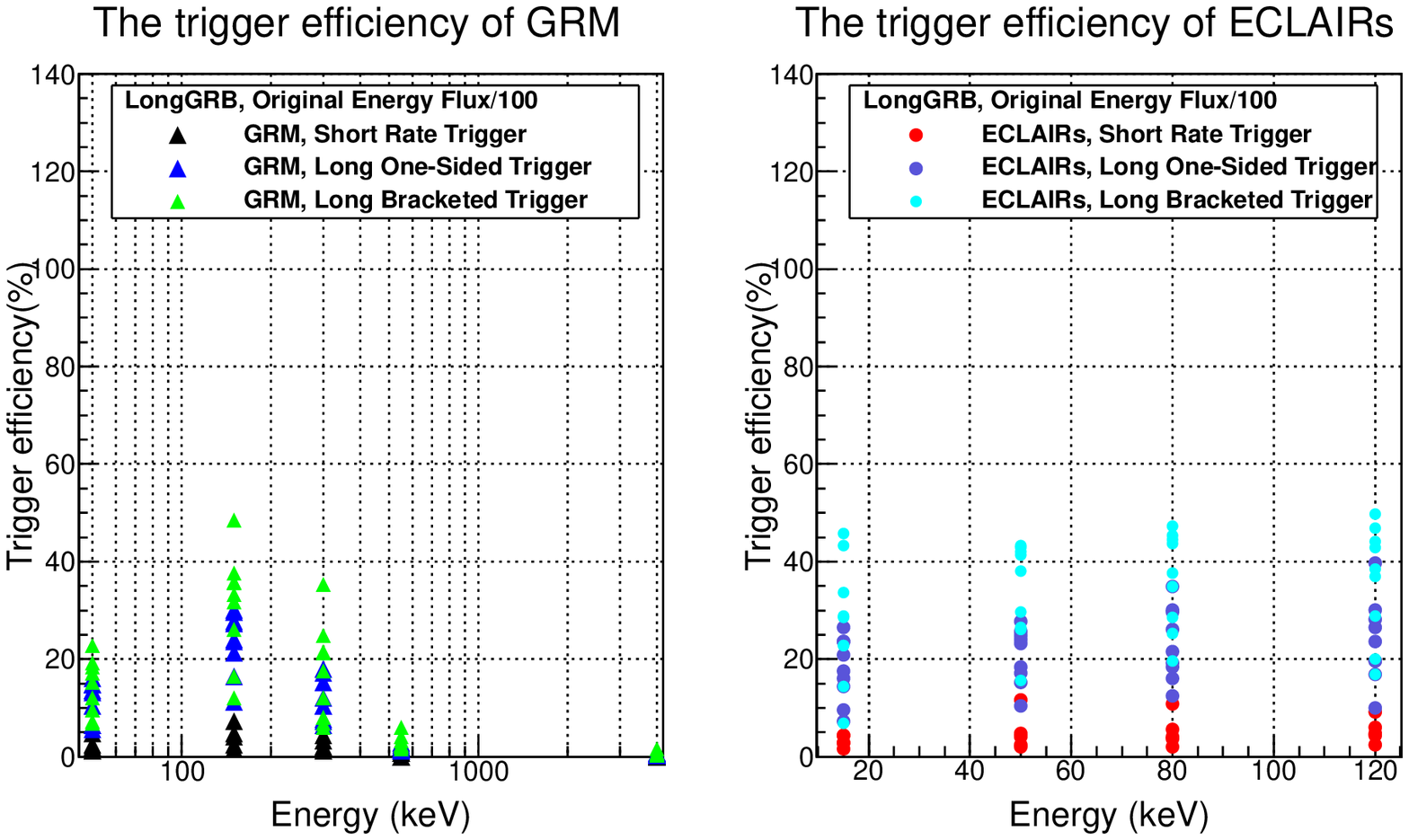}	
\caption{The efficiencies of GRM and ECLAIRs for bright (top) and dark (bottom) long GRBs in different energy ranges.}
\label{Fig:eff-energy-long}
\end{figure}

\begin{figure}[!ht]
\centering
\includegraphics[width=0.7\textwidth]{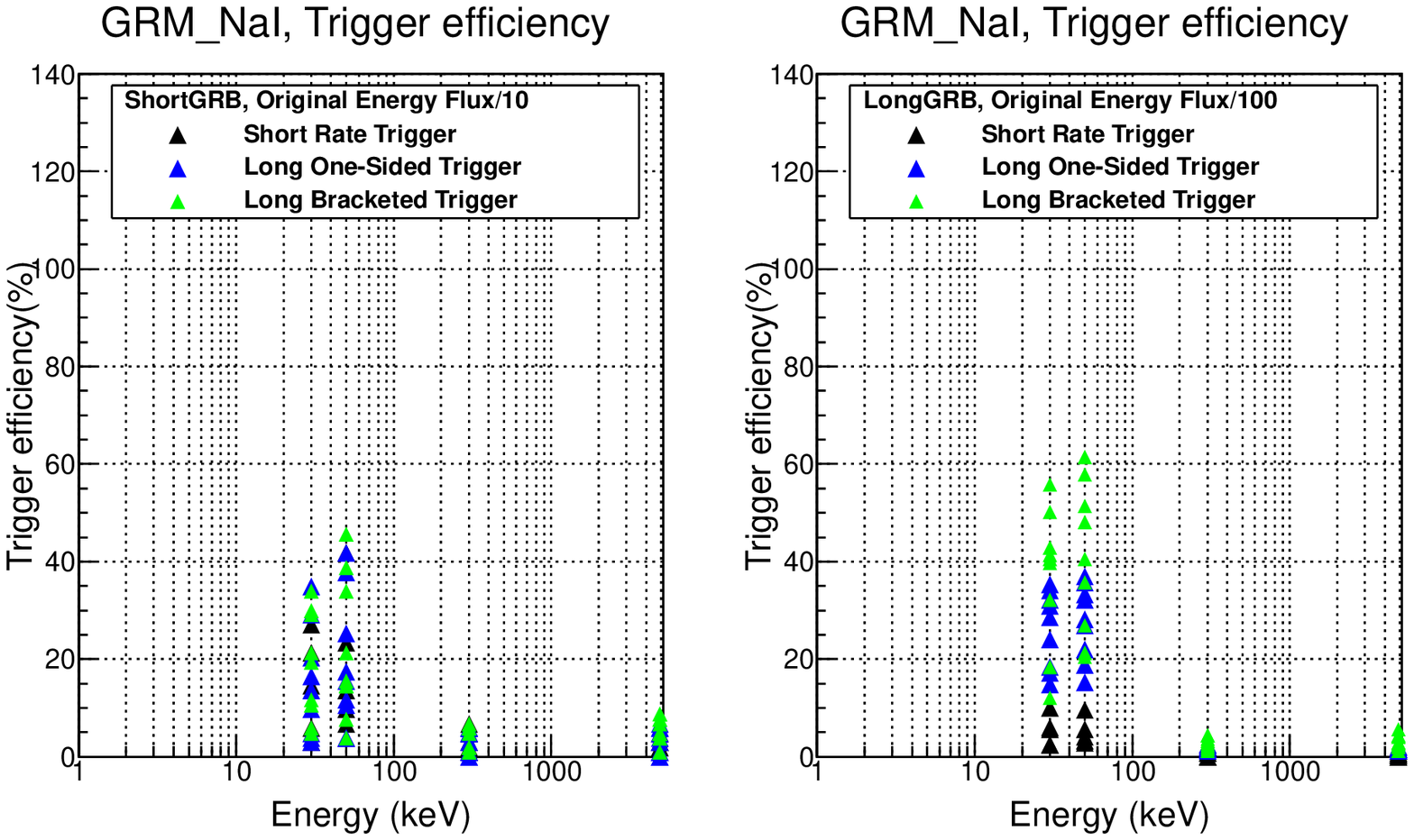}	
\caption{The efficiencies of GRM for dark short (left) and dark long (right) GRBs in larger energy ranges
(30 keV, 50 keV, 300 keV and 5000 keV indicate the energy ranges 30-150 keV, 50-300 keV, 300-1000 keV and 300-5000 keV, respectively).}
\label{Fig:eff-largerEn}
\end{figure}

The variations of the trigger efficiency for short and long GRBs with energy ranges are shown in Fig.~\ref{Fig:eff-energy-short} and ~\ref{Fig:eff-energy-long}, respectively.
From the maximum efficiency in each energy range, we can conclude that 50-150 keV and 150-300 keV are the most sensitive energy ranges for GRM.
The efficiency for dark GRBs is obviously lower than that for bright GRBs which may be due to the narrowness of the energy ranges.
Accordingly, we combine different energy ranges to form a larger one. And the corresponding results are presented in Fig.~\ref{Fig:eff-largerEn}
which shows that even though the efficiency can be somewhat raised by enlarging the trigger energy range, the efficiency of dark GRBs is still much lower.
And that the energy ranges below 300 keV are still the most sensitive. For ECLAIRs, the highest trigger efficiency of all kinds of GRBs in every energy range is very similar.

\begin{figure}
\centering
\includegraphics[height=0.25\textheight, width=0.7\textwidth]{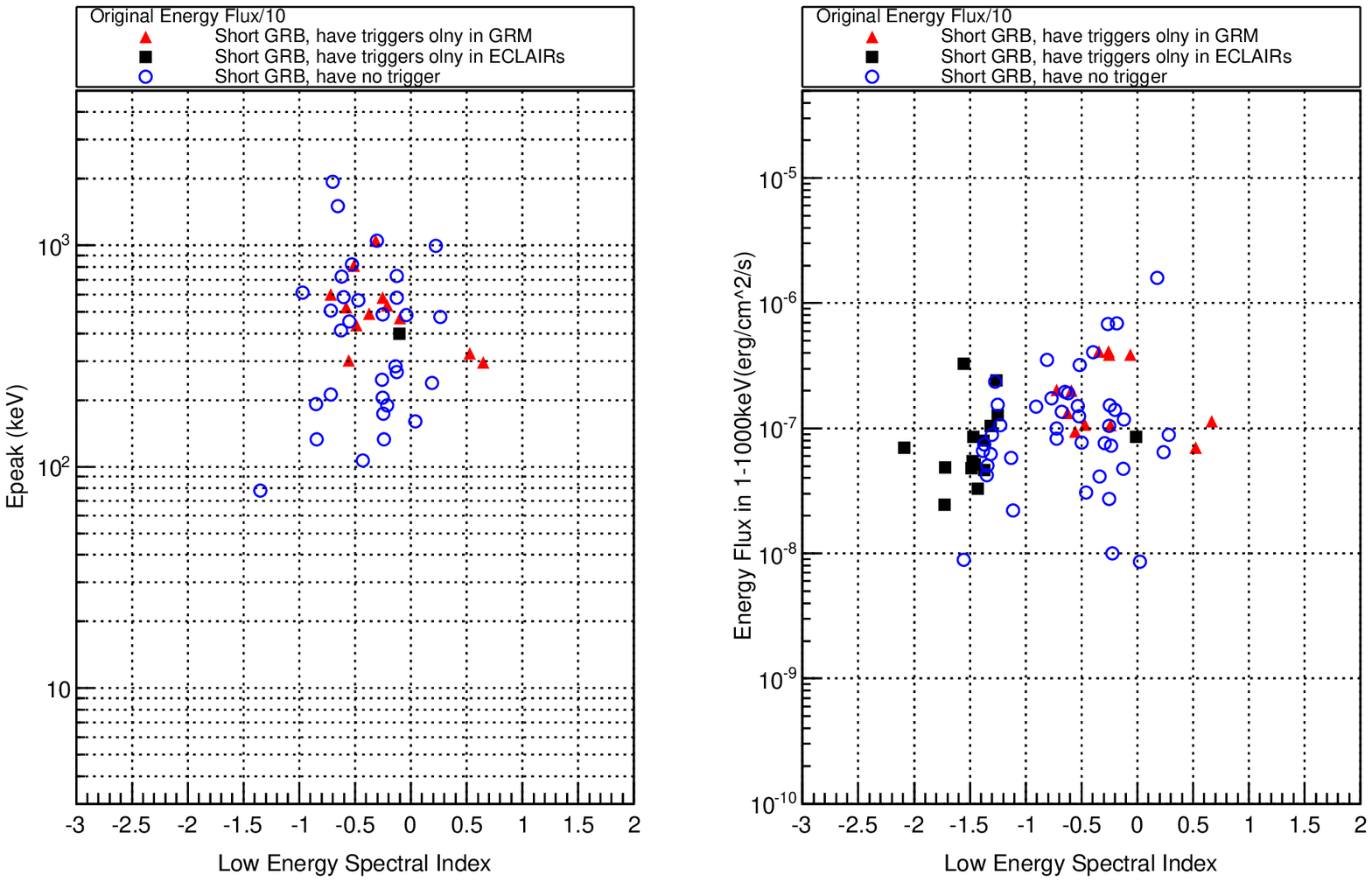}	
\caption{The distributions of the dark short GRBs which have triggers in GRM or in ECLAIRs.}
\label{Fig:shortflux01}
\end{figure}

\begin{figure}
\centering
\includegraphics[height=0.25\textheight, width=0.7\textwidth]{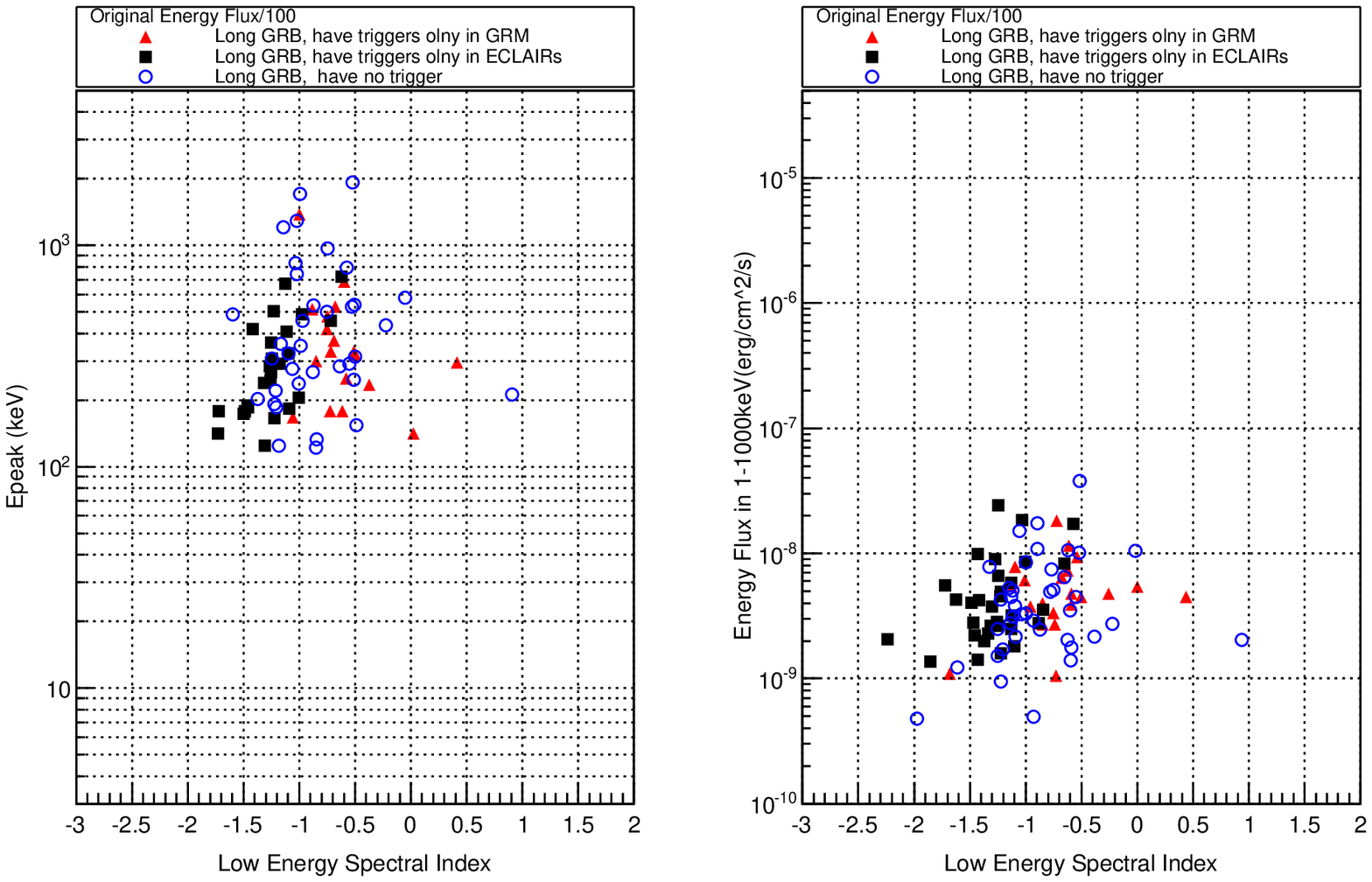}	
\caption{The distributions of the dark long GRBs which have triggers in GRM or in ECLAIRs.}
\label{Fig:longflux001}
\end{figure}

Almost all of the bright GRBs can be triggered in both GRM and ECLAIRs, especially the bright long GRBs.
Therefore, we investigate the characteristics of GRBs which have triggers in GRM or ECLAIRs using dark GRBs.
We extract three kinds of GRBs which have triggers only in GRM, only in ECLAIRs and which have triggers in neither one, respectively. 
And we denote these three kinds of GRBs by GRM-GRBs, ECLAIRs-GRBs and NO-GRBs for convenience, respectively.
The numbers of these GRBs are shown in Table~\ref{Tab:grbtrigger}.
And their distributions on Low energy spectral index vs. $E_{\rm peak}$ and Low energy spectral index vs.
Energy flux are shown in Fig.~\ref{Fig:shortflux01} and~\ref{Fig:longflux001} in which the three kinds of GRBs 
are denoted by solid triangles, solid squares and hollow circles, respectively.
Some of the GRBs have no the parameter of $E_{\rm peak}$, therefore, the data in the left panels of Fig.~\ref{Fig:shortflux01} and Fig.~\ref{Fig:longflux001}
are less than those in the right panels. 

As the flux decrease, the characteristics of GRM-GRBs, ECLAIRs-GRBs and NO-GRBs are as follows:
1) The numbers of GRM-GRBs, ECLAIRs-GRBs and NO-GRBs increase gradually before the flux decreases too much;
2) The low energy spectral index of GRM-GRBs are larger than that of ECLAIRs-GRBs;
3) The proportion of the GRBs with single power-law spectrum in ECLAIRs-GRBs is much larger than that in GRM-GRBs.
2) and 3) imply that GRM is more sensitive to hard GRBs comparing with ECLAIRs, and that GRM will play a very important role in measuring the $E_{\rm peak}$， especially the $E_{\rm peak}$ of short GRBs.

\begin{table}
\begin{center}
\caption{The numbers of GRBs, which have triggers in GRM or ECLAIRs, when the GRBs have 1/n of flux of the bright GRBs.
(GRM: have triggers only in GRM.	ECLAIRs: have triggers only in ECLAIRs.	No: have triggers neither in GRM nor in ECLAIRs.	Yes: have triggers in GRM or in ECLAIRs.
The values in brackets are the numbers of the GRBs with spectrum of single power-law.)}
\label{Tab:grbtrigger}
\small 
\begin{tabular}{cllllll}
\hline
\hline
GRB Type    &		&Trigger Type    &Flux     &Flux/5  &Flux/10     &Flux/100\\
\hline\noalign{\smallskip}
            &		&GRM              &3        &9      &12(0)             &0\\

Short GRB   &		&ECLAIRs          &1        &9      &15(14)             &2\\

            &		&No               &5        &17     &42             &100\\

            &		&Yes              &98       &86     &61             &3\\
\hline\noalign{\smallskip}
            &		&GRM              &0        &0      &0              &21(1) \\

Long GRB    &		&ECLAIRs          &0        &0      &0              &32(3)\\

            &		&No               &0        &0      &0              &40\\

            &		&Yes              &249      &249     &249          &209\\
\hline
\hline

\end{tabular}%
\end{center}
\end{table}

\section{The solar flare triggers}
\label{sec:sf}

Solar flares (Lin~\cite{Lin2011}; Fletcher et al.~\cite{Fletcher2011}) are the most powerful explosions in the solar system.    
Out of the 8021 triggers of BATSE, the proportions of GRBs and solar flares are 33.7\% and 14.8\%\footnote{ftp://umbra.nascom.nasa.gov/pub/batse/batse0/burst\_trigger.lst}, respectively.
There are so many triggers induced by solar flares on BATSE that it is necessary to estimate the impact of flares on GRM which is a scintillator detector.
SVOM will point close the anti-solar direction during a large fraction of the orbit. So, we can predict that the probability of the triggers caused by solar flares
in GRM will be much lower than that in BATSE. BATSE distinguish flares from GRBs mainly by location and spectrum combined with the observations of
GOES (Mallozzi et al.~\cite{Mallozzi1993}). The pointing strategy of SVOM makes ECLAIRs unable to image and localize solar flares.
Thus, the distinction of solar flares can only rely on the performance of GRM or the combinations with the observations of other instruments.
In this section, we will study how to distinguish the solar flares from GRBs using the data of GRM\_NaI and GRM\_CsI.

We use the solar flares observed by RHESSI as the sample. Getting rid of the events which have no position information or whose positions are zero or
whose maximum energy is below 6 keV, there are totally 57388 solar flares from 12 February 2002 to 27 May 2012 according to the data
on the web site\footnote{http://hesperia.gsfc.nasa.gov/hessidata/dbase/}. On the one hand, we statistic the distribution of these solar flares; on the other hand,
we do simulations with one typical flare. We estimate the average number of flares which can be detected by GRM per year and investigate the method to distinguish solar flares from GRBs via these two aspects of work.

The distributions of the flares on the highest energy band and on the peak count per second in energy range 12-25 keV are shown
in Table~\ref{Tab:RHESSIsf}. Most of the solar flares are soft and are not very intense. The flares with highest energy $\geq$50 keV is only 2.3\%, and those
with peak counts $\geq$400 in 12-25 keV is 11.8\%. When photons inject from behind, the efficiency of GRM\_NaI for the 100 keV photons is 0.85\%,
and for the 300 keV and 800 keV photons are 4.4\% and 8.5\%, respectively. Accordingly, we can expect that only the solar flares with high energy and large peak counts can be detected by GRM.

\begin{table}
\begin{center}
\caption{The distributions of the solar flares observed by RHESSI on the maximum energy and on the peak count in the energy range 12-25 keV.}
\label{Tab:RHESSIsf}
\small 
\begin{tabular}{cll|cll}
\hline
\hline
$\geq$E0(keV)            &N(Flare)      &Proportion      &$\geq$Peak Count(counts/s)      &N(Flare)    &Proportion\\
\hline
                12	             &57388	                 &1                  &400             &6763             &0.1179\\
                25	             &12550	                 &0.21869            &800             &3719             &0.0648\\
                50	             &1333	                 &0.02323            &1000            &2958             &0.0515\\
                100	             &221	                 &0.00385            &10000           &50               &0.0009\\
                300	             &75	                 &0.00131            &-               &-                   &-\\
                800	             &11	                 &0.00019            &-               &-                   &-\\
                7000	         &4	                     &0.00007            &-               &-                   &-\\
\hline
\hline
\end{tabular}%
\end{center}
\end{table}

Solar flare SOL2002-07-23(X4.8), whose spectrum and light curves can be found in (Lin~\cite{Lin2011}) and on the web
site \footnote{http://hesperia.gsfc.nasa.gov/hessidata/dbase/}, is a very typical intense flare.
For this flare, the measurements span almost 4 orders of magnitude in photon energy and more than 12 orders in flux. SOL2002-07-23 lasts about one hour
and its peak count in energy range 12-25 keV is about 42341 counts/s.
We adopt its spectrum and light curve in energy range 25-7000 keV.

\begin{figure}
\centering
\includegraphics[width=14cm]{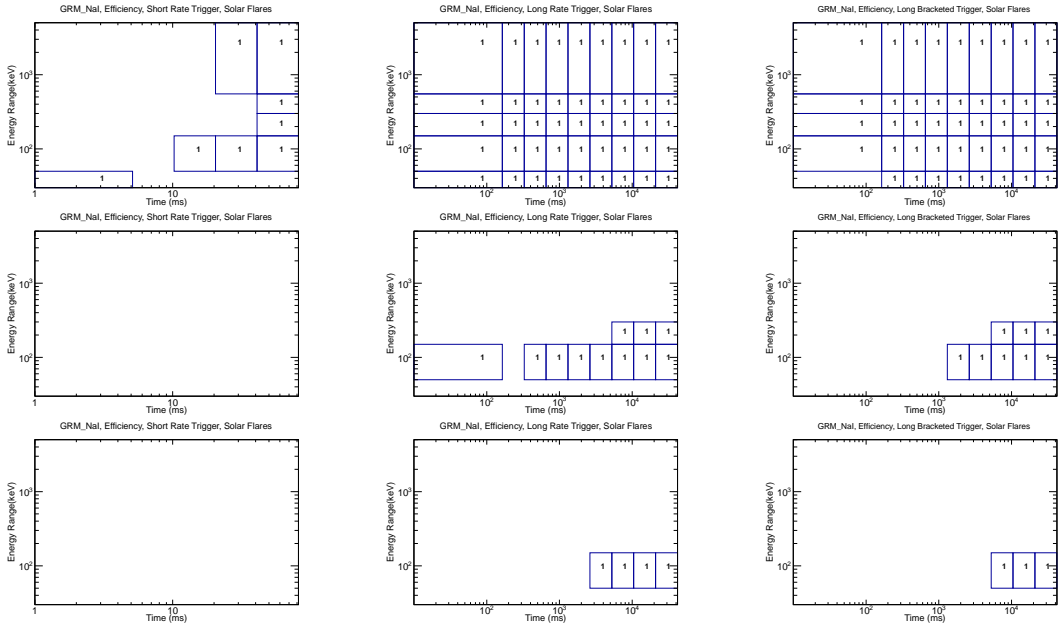}	
\caption{The triggers of SOL2002-07-23 in GRM\_NaI. From top to bottom, the corresponding classifications are X4.8, M2.4 and M1, respectively
(1 in the figure means that the flare has trigger in the corresponding energy range-time scale combination and empty means no trigger).}  
\label{Fig:sfnai}
\end{figure}

\begin{figure}
\centering
\includegraphics[width=14cm]{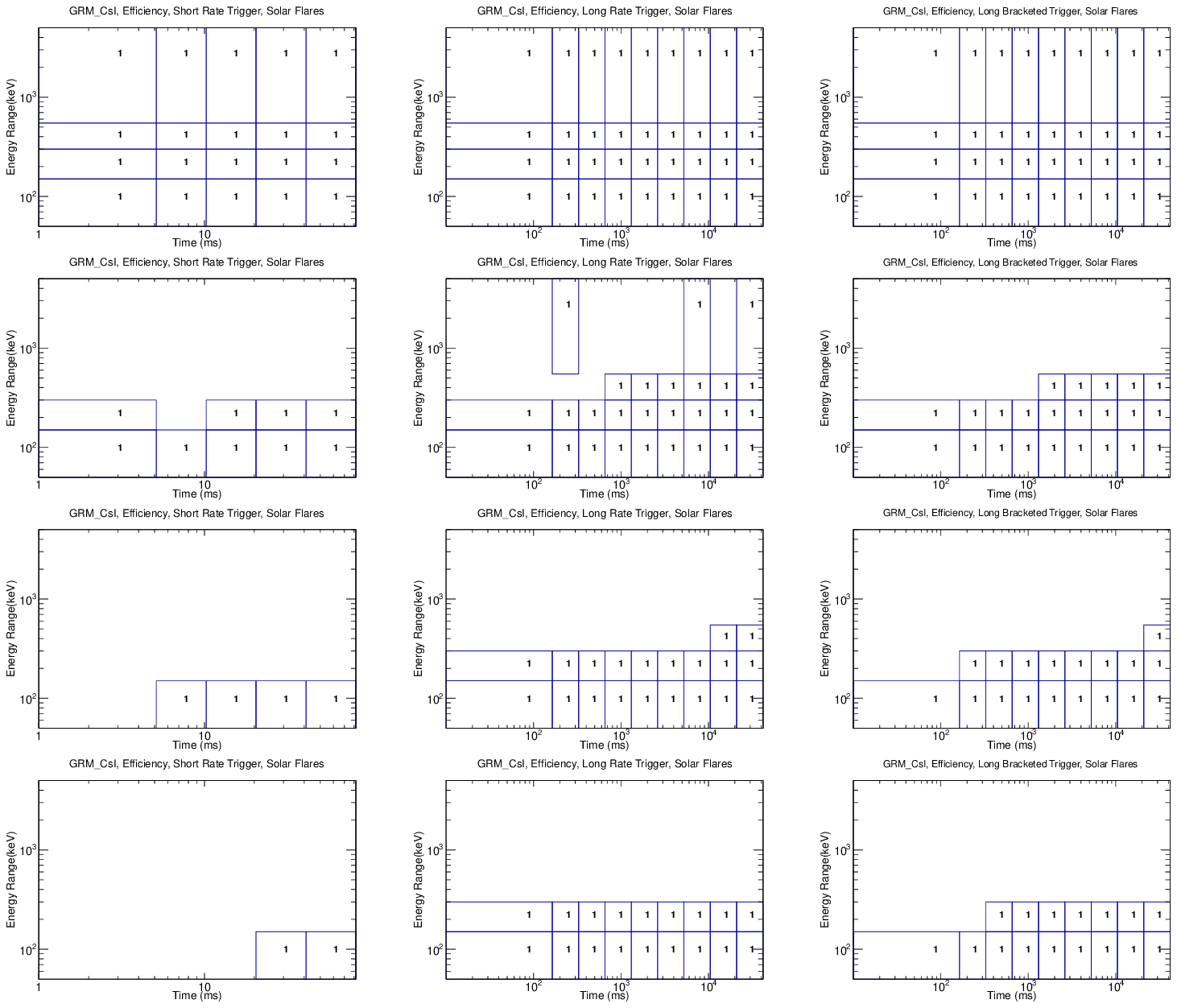}	
\caption{The triggers of SOL2002-07-23 in GRM\_CsI. From top to bottom, the corresponding classifications are X4.8, M2.4, M1 and C4.8, respectively.}
\label{Fig:sfcsi}
\end{figure}

Like what we did for GRBs, we decrease the flux of SOL2002-07-23 keeping the shapes of spectrum and light curve to obtain the flares of M2.4, M1 and C4.8.
And for different classifications of flare, we get the trigger results in GRM\_NaI and GRM\_CsI as shown in Fig.~\ref{Fig:sfnai} and~\ref{Fig:sfcsi}.
The flare of X4.8 has triggers in every energy range of GRM\_NaI and GRM\_CsI, though it has almost no trigger in GRM\_NaI on the
time scale less than 20 ms.
For the flare of M2.4, there is no trigger on short time scales; the main trigger energy range of GRM\_NaI is 50-150 keV;
and there are very few triggers in energy range 150-300 keV on longer time scales ($\geq$10 sec). In GRM\_NaI, there is not any trigger for the flare of C4.8 whose peak count is 423 counts/s.  
Assuming that only the flares, which have energy emissions above 50 keV and peak counts more than 400 counts/s, have triggers in GRM\_NaI.
And then 187 (about 0.3\%) out of the solar flares observed by RHESSI will have triggers in GRM. Namely, there are less than 20 solar flares per year which can cause false triggers in GRM\_NaI.

\begin{figure}
\centering
\includegraphics[width=0.7\textwidth]{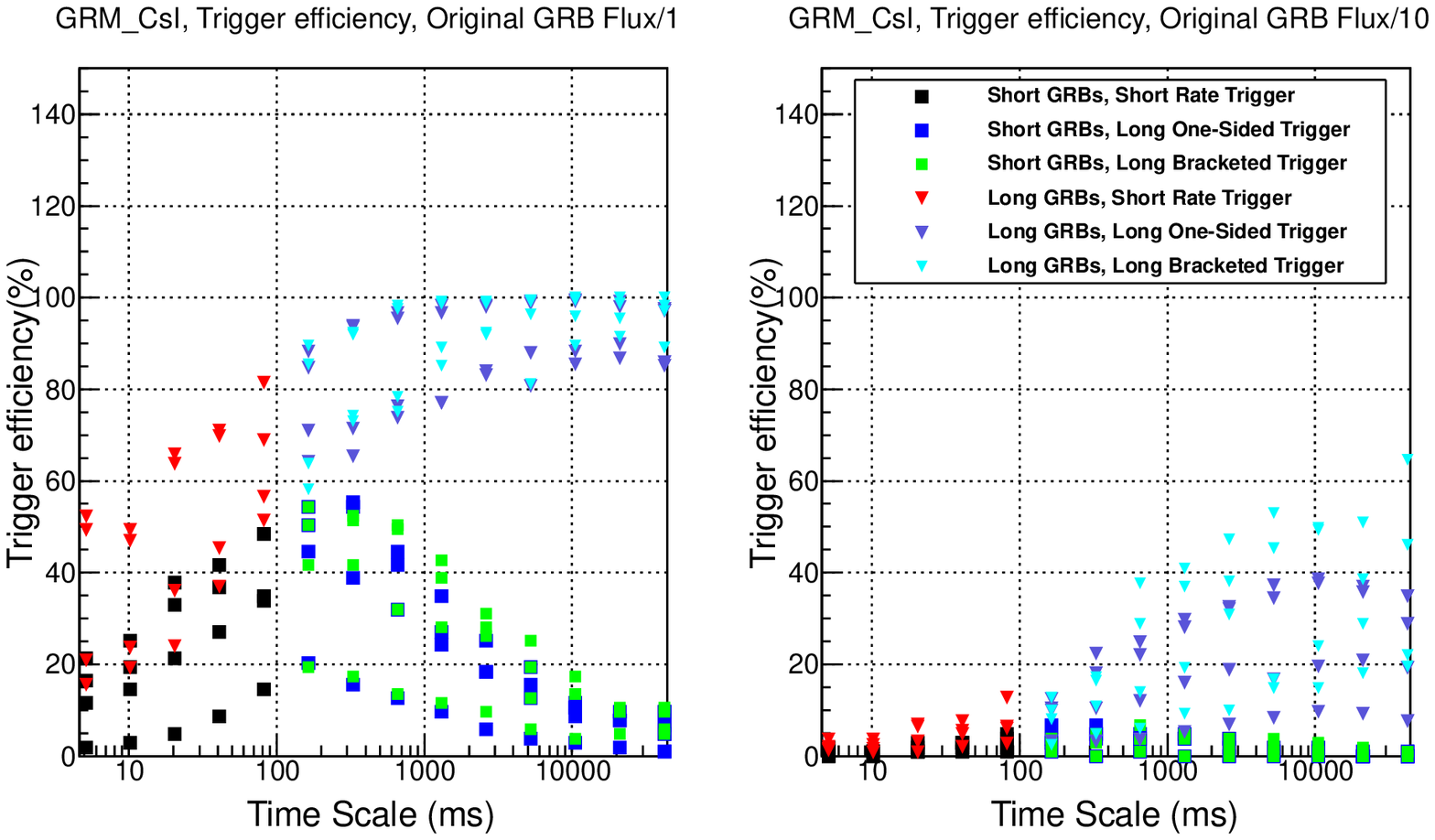}	
\includegraphics[width=0.7\textwidth]{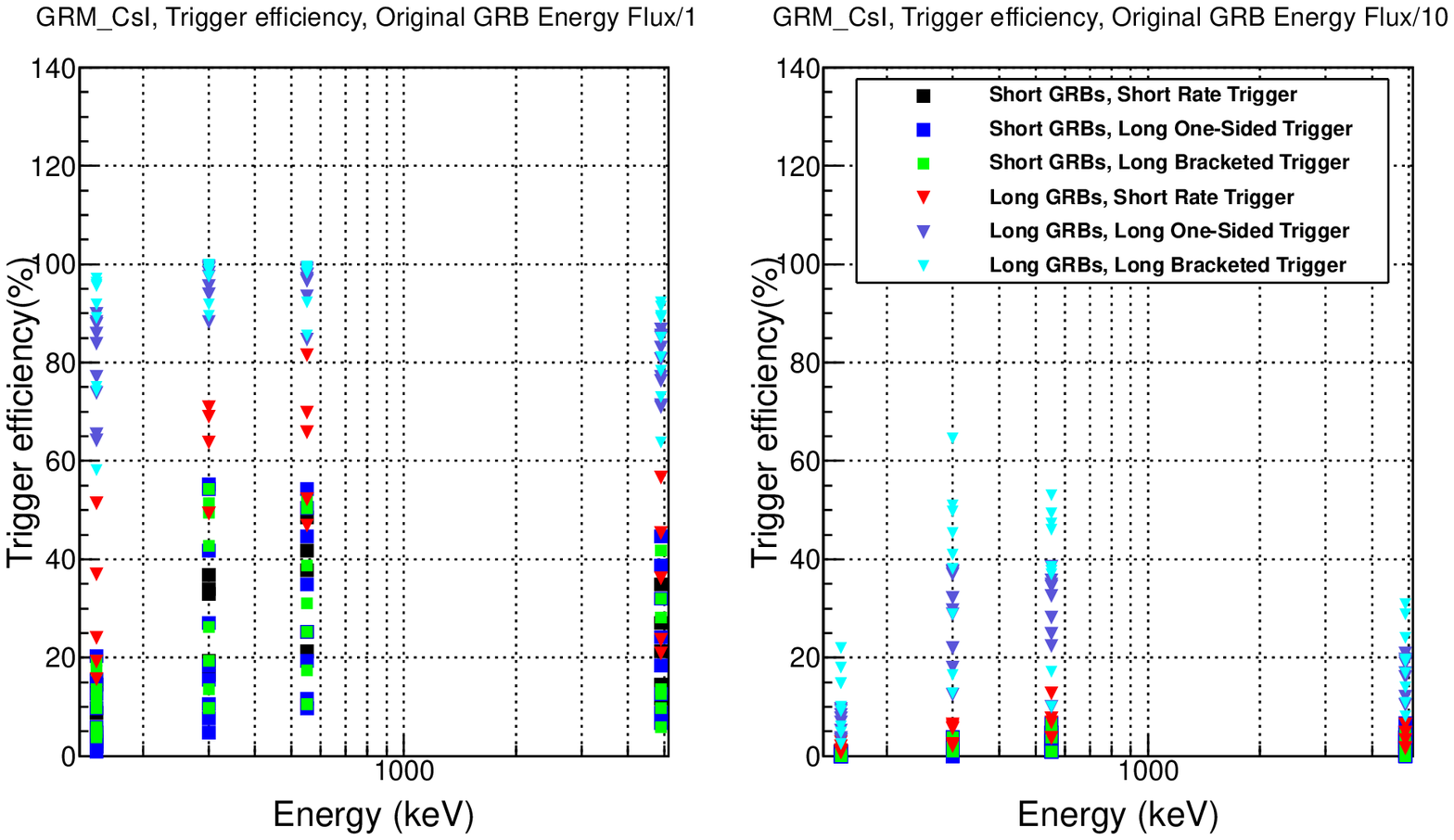}	
\caption{The efficiencies of GRM\_CsI for bright (left) and dark (right) GRBs on different time scales (top) and in different energy ranges(bottom,
150 keV, 300 keV, 550 keV and 5000 keV indicate the energy ranges 50-150 keV, 150-300 keV, 300-550 keV and 550-5000 keV, respectively).
}
\label{Fig:eff-csi-grb}
\end{figure}

We described the trigger efficiency of GRBs in GRM\_NaI in Fig.~\ref{Fig:eff-energy-short} and \ref{Fig:eff-energy-long}. Comparing with solar flares,
GRM\_NaI is very sensitive to GRBs in the energy range 150-300 keV in addition to 50-150 keV.
As shown in Fig.~\ref{Fig:sfcsi}, 50-150 keV and 150-300 keV are two main trigger energy ranges of GRM\_CsI for solar flares, especially 50-150 keV.
Fig.~\ref{Fig:eff-csi-grb} shows that GRM\_CsI is not sensitive to short GRBs (the maximum efficiency is 55.3\% for bright short GRBs).
And the efficiency for bright long GRBs can be more than 90\% on the time scales larger than 160 ms.	
When the flux of GRBs decreases, the energy range 50-150 keV of GRM\_CsI is not sensitive anymore.

For GRBs, the trigger efficiency in GRM\_NaI is often obviously higher than that in GRM\_CsI for the same energy range and time scale combination.
And for solar flares, it is converse. Comparing the distributions of energy ranges in which the GRBs and solar flares have triggers with relatively high efficiency, 
we can conclude that:
1) the sources not triggered in GRM\_CsI are not likely to be solar flares;
2) the sources triggered in GRM\_CsI and not triggered in GRM\_NaI are not GRBs in FOV;
3) the sources triggered outside 50-150 keV in GRM\_NaI are not likely to be solar flares.

\section{Summary}
\label{sec:sum}
In this work, three on-board counting rate trigger algorithms of GRM are investigated. And the prime difference between different algorithms is the method to calculate the background. They have their own advantages and disadvantages and complement each other. Short rate trigger algorithm, which is fast but has high threshold and rough
calculation of background, is suitable to short GRB triggers but not to dark GRB triggers. Long one-sided trigger algorithm, which can remove the trend of background and has lower threshold, is applicable to long GRB triggers. Long bracketed trigger algorithm, which can remove the trend of background most accurately,
has advantage to detecte long GRBs especially to long dark GRBs, but it takes longer than other two algorithms.
We can use them simultaneously in order to increase trigger efficiency and to detecte GRBs as early as possible.
The computations of trigger efficiency based on the given GRB sample show that 50-150 keV and 150-300 keV are the most sensitive trigger energy ranges of GRM
and that a few seconds and several hundred milliseconds are the sensitive time scales for long GRBs and shor GRBs, respectively.
We applied the same trigger algorithms to ECLAIRs, and investigated the trigger characteristics of GRM and ECLAIRs. 
In addition, we find that the solar flares can be disguished from GRBs by analyzing the distribution of triggers in energy ranges of GRM\_NaI and GRM\_CsI. And less than 20 solar flares per year, on average, can cause triggers in GRM\_NaI according to the statistics of the solar flares observed by RHESSI from 2002 to 2012 and the simulations using the typical SOL2002-07-23.

We used a simplification during the simulations.
We keep the GRBs in the center of the FOV and keep the photons perpendicular to the detector plane which is the ideal case.
And with this simplification, the detection efficiency of the instruments is usually the highest which might make
the trigger efficiency higher than that it should be. 
The investigation with GRBs in various positions in the FOV needs more further work.

According to the observations of BATSE, discrete source fluctuations, Soft Gamma Repeaters(SGRs) and the magnetospheric events
which mainly are the electron precipitation events (Mallozzi et al.~\cite{Mallozzi1993}) can also cause many triggers.
The triggers occurred as a transient source emerges from behind the Earth can be easily classified since it appears as occultation steps at the predictable times.
And the SGRs can be identified primarily by their soft spectra and typical short durations (about 0.1 sec) (Mallozzi et al.~\cite{Mallozzi1993}; Kouveliotou et al.~\cite{Kouveliotou1992}). However, the key to identify the triggers by the electron precipitation events and the source flares is the localization capability of instrument from the experiences of BATSE (Horack et al.~\cite{Horack1991}; Meegan et al.~\cite{Meegan1993}). GRM consisting of two identical detectors which point to the same direction has no localization capability. And ECLAIRs has image and localization capabilities. Accordingly, GRM can identify the GRB triggers and reject the false triggers more effectively taking full advantage of the trigger information of ECLAIRs. Therefor, the investigation of the cooperations between GRM and ECLAIRs on GRB triggers is another important work which will be proceed in the following work based on the work described in this paper.

\label{lastpage}

\begin{thebibliography}{37}
\providecommand{\natexlab}[1]{#1}
\providecommand{\selectlanguage}[1]{\relax}

\bibitem[{{Boyle} et~al.(2000){Boyle}, {Shanks}, {Croom}
  et~al.}]{Boyle+etal+2000}
{Boyle}, B.~J., {Shanks}, T., {Croom}, S.~M., et~al. 2000, \mnras, 317, 1014

\bibitem[{{Casali} et~al.(2007){Casali}, {Adamson}, {Alves de Oliveira}
  et~al.}]{Casali+etal+2007}
{Casali}, M., {Adamson}, A., {Alves de Oliveira}, C., et~al. 2007, \aap, 467,
  777

\bibitem[{{Chiu} et~al.(2007){Chiu}, {Richards}, {Hewett}, \&
  {Maddox}}]{Chiu+etal+2007}
{Chiu}, K., {Richards}, G.~T., {Hewett}, P.~C., \& {Maddox}, N. 2007, \mnras,
  375, 1180

\bibitem[{{Constantin} et~al.(2002){Constantin}, {Shields}, {Hamann}, {Foltz},
  \& {Chaffee}}]{Constantin+etal+2002}
{Constantin}, A., {Shields}, J.~C., {Hamann}, F., {Foltz}, C.~B., \& {Chaffee},
  F.~H. 2002, \apj, 565, 50

\bibitem[{{Croom} et~al.(2004){Croom}, {Smith}, {Boyle}
  et~al.}]{Croom+etal+2004}
{Croom}, S.~M., {Smith}, R.~J., {Boyle}, B.~J., et~al. 2004, \mnras, 349, 1397

\bibitem[{{Fan} et~al.(2001{\natexlab{a}}){Fan}, {Narayanan}, {Lupton}
  et~al.}]{Fan+etal+2001b}
{Fan}, X., {Narayanan}, V.~K., {Lupton}, R.~H., et~al. 2001{\natexlab{a}}, \aj,
  122, 2833

\bibitem[{{Fan} et~al.(2001{\natexlab{b}}){Fan}, {Strauss}, {Schneider}
  et~al.}]{Fan+etal+2001a}
{Fan}, X., {Strauss}, M.~A., {Schneider}, D.~P., et~al. 2001{\natexlab{b}},
  \aj, 121, 54

\bibitem[{{Hambly} et~al.(2008){Hambly}, {Collins}, {Cross}
  et~al.}]{Hambly+etal+2008}
{Hambly}, N.~C., {Collins}, R.~S., {Cross}, N.~J.~G., et~al. 2008, \mnras, 384,
  637

\bibitem[{{Hewett} et~al.(2006){Hewett}, {Warren}, {Leggett}, \&
  {Hodgkin}}]{Hewett+etal+2006}
{Hewett}, P.~C., {Warren}, S.~J., {Leggett}, S.~K., \& {Hodgkin}, S.~T. 2006,
  \mnras, 367, 454

\bibitem[{{Hu} et~al.(2008){Hu}, {Wang}, {Ho} et~al.}]{Hu+etal+2008}
{Hu}, C., {Wang}, J.-M., {Ho}, L.~C., et~al. 2008, \apj, 687, 78

\bibitem[{{Kong} et~al.(2006){Kong}, {Wu}, {Wang}, \& {Han}}]{Kong+etal+2006}
{Kong}, M.-Z., {Wu}, X.-B., {Wang}, R., \& {Han}, J.-L. 2006, \cjaa, 6, 396

\bibitem[{{Lawrence} et~al.(2007){Lawrence}, {Warren}, {Almaini}
  et~al.}]{Lawrence+etal+2007}
{Lawrence}, A., {Warren}, S.~J., {Almaini}, O., et~al. 2007, \mnras, 379, 1599

\bibitem[{{Maddox} et~al.(2008){Maddox}, {Hewett}, {Warren}, \&
  {Croom}}]{Maddox+etal+2008}
{Maddox}, N., {Hewett}, P.~C., {Warren}, S.~J., \& {Croom}, S.~M. 2008, \mnras,
  386, 1605

\bibitem[{{Markwardt}(2009)}]{Markwardt+2009}
{Markwardt}, C.~B. 2009, in Astronomical Data Analysis Software and Systems
  XVIII, \emph{Astronomical Society of the Pacific Conference Series}, vol.
  411, edited by D.~A. {Bohlender}, D.~{Durand}, \& P.~{Dowler}, 251

\bibitem[{{Richards} et~al.(2002){Richards}, {Fan}, {Newberg}
  et~al.}]{Richards+etal+2002}
{Richards}, G.~T., {Fan}, X., {Newberg}, H.~J., et~al. 2002, \aj, 123, 2945

\bibitem[{{Richards} et~al.(2009){Richards}, {Myers}, {Gray}
  et~al.}]{Richards+etal+2009}
{Richards}, G.~T., {Myers}, A.~D., {Gray}, A.~G., et~al. 2009, \apjs, 180, 67

\bibitem[{{Schlegel} et~al.(1998){Schlegel}, {Finkbeiner}, \&
  {Davis}}]{Schlegel+etal+1998}
{Schlegel}, D.~J., {Finkbeiner}, D.~P., \& {Davis}, M. 1998, \apj, 500, 525

\bibitem[{{Schmidt}(1963)}]{Schmidt+1963}
{Schmidt}, M. 1963, \nat, 197, 1040

\bibitem[{{Schneider} et~al.(2010){Schneider}, {Richards}, {Hall}
  et~al.}]{Schneider+etal+2010}
{Schneider}, D.~P., {Richards}, G.~T., {Hall}, P.~B., et~al. 2010, \aj, 139,
  2360

\bibitem[{{Shen} et~al.(2011){Shen}, {Richards}, {Strauss}
  et~al.}]{Shen+etal+2011}
{Shen}, Y., {Richards}, G.~T., {Strauss}, M.~A., et~al. 2011, \apjs, 194, 45

\bibitem[{{Smith} et~al.(1994){Smith}, {Djorgovski}, {Thompson}
  et~al.}]{Smith+etal+1994}
{Smith}, J.~D., {Djorgovski}, S., {Thompson}, D., et~al. 1994, \aj, 108, 1147

\bibitem[{{Smith} et~al.(2005){Smith}, {Croom}, {Boyle}
  et~al.}]{Smith+etal+2005}
{Smith}, R.~J., {Croom}, S.~M., {Boyle}, B.~J., et~al. 2005, \mnras, 359, 57

\bibitem[{{Su} et~al.(1998){Su}, {Cui}, {Wang}, \& {Yao}}]{Su+etal+1998}
{Su}, D.~Q., {Cui}, X., {Wang}, Y., \& {Yao}, Z. 1998, in Society of
  Photo-Optical Instrumentation Engineers (SPIE) Conference Series,
  \emph{Society of Photo-Optical Instrumentation Engineers (SPIE) Conference
  Series}, vol. 3352, edited by L.~M. {Stepp}, 76--90

\bibitem[{{Tsuzuki} et~al.(2006){Tsuzuki}, {Kawara}, {Yoshii}
  et~al.}]{Tsuzuki+etal+2006}
{Tsuzuki}, Y., {Kawara}, K., {Yoshii}, Y., et~al. 2006, \apj, 650, 57

\bibitem[{{Vestergaard} \& {Peterson}(2006)}]{Vestergaard+Peterson+2006}
{Vestergaard}, M., \& {Peterson}, B.~M. 2006, \apj, 641, 689

\bibitem[{{Vestergaard} \& {Wilkes}(2001)}]{Vestergaard+Wilkes+2001}
{Vestergaard}, M., \& {Wilkes}, B.~J. 2001, \apjs, 134, 1

\bibitem[{{Warren} et~al.(2000){Warren}, {Hewett}, \&
  {Foltz}}]{Warren+etal+2000}
{Warren}, S.~J., {Hewett}, P.~C., \& {Foltz}, C.~B. 2000, \mnras, 312, 827

\bibitem[{{Wright} et~al.(2010){Wright}, {Eisenhardt}, {Mainzer}
  et~al.}]{Wright+etal+2010}
{Wright}, E.~L., {Eisenhardt}, P.~R.~M., {Mainzer}, A.~K., et~al. 2010, \aj,
  140, 1868

\bibitem[{{Wu} et~al.(2010{\natexlab{a}}){Wu}, {Chen}, {Jia}
  et~al.}]{Wu+etal+2010a}
{Wu}, X.-B., {Chen}, Z.-Y., {Jia}, Z.-D., et~al. 2010{\natexlab{a}}, Research
  in Astronomy and Astrophysics, 10, 737

\bibitem[{{Wu} \& {for the LAMOST Extragalactic Survey Team}(2011)}]{Wu+2011}
{Wu}, X.-B., \& {for the LAMOST Extragalactic Survey Team} 2011, ArXiv e-prints

\bibitem[{{Wu} et~al.(2012){Wu}, {Hao}, {Jia}, {Zhang}, \&
  {Peng}}]{Wu+etal+2012}
{Wu}, X.-B., {Hao}, G., {Jia}, Z., {Zhang}, Y., \& {Peng}, N. 2012, \aj, 144,
  49

\bibitem[{{Wu} \& {Jia}(2010)}]{Wu+Jia+2010}
{Wu}, X.-B., \& {Jia}, Z. 2010, \mnras, 406, 1583

\bibitem[{{Wu} et~al.(2010{\natexlab{b}}){Wu}, {Jia}, {Chen}
  et~al.}]{Wu+etal+2010b}
{Wu}, X.-B., {Jia}, Z.-D., {Chen}, Z.-Y., et~al. 2010{\natexlab{b}}, Research
  in Astronomy and Astrophysics, 10, 745

\bibitem[{{Wu} et~al.(2011){Wu}, {Wang}, {Schmidt} et~al.}]{Wu+etal+2011}
{Wu}, X.-B., {Wang}, R., {Schmidt}, K.~B., et~al. 2011, \aj, 142, 78

\bibitem[{{Wu} et~al.(2004){Wu}, {Zhang}, \& {Zhou}}]{Wu+etal+2004}
{Wu}, X.-B., {Zhang}, W., \& {Zhou}, X. 2004, \cjaa, 4, 17

\bibitem[{{York} et~al.(2000){York}, {Adelman}, {Anderson}
  et~al.}]{York+etal+2000}
{York}, D.~G., {Adelman}, J., {Anderson}, J.~E., Jr., et~al. 2000, \aj, 120,
  1579

\bibitem[{{Zhao} et~al.(2012){Zhao}, {Zhao}, {Chu}, {Jing}, \&
  {Deng}}]{Zhao+etal+2012}
{Zhao}, G., {Zhao}, Y.-H., {Chu}, Y.-Q., {Jing}, Y.-P., \& {Deng}, L.-C. 2012,
  Research in Astronomy and Astrophysics, 12, 723

\end{thebibliography}


\begin{thebibliography}{99}

\bibitem[2008]{Basa2008} Basa S., Wei J. et al., 2008, SF2A. Conf., 161.
\bibitem[2009]{Dong2009} Dong Y.-W., Wu B.-B. et al., 2009, Sci. China. Ser. G-Phys. Mech. Astron., 52, 1
\bibitem[2001]{Fenimore2001} Fenimore E.E. and Galassi M., 2001, ESO Symposia: Gamma-ray Burst in Afterglow, 393
\bibitem[2004]{Fenimore2004} Fenimore E. E., McLean K., Palmer D. et al., 2004, Baltic Astronomy, 13, 301
\bibitem[2001]{FenimoreD2001} Fenimore E. E., Palmer D., Galassi M. et al., 2001, Gamma-Ray Burst and Afterglow Astronomy.
\bibitem[2011]{Fletcher2011} Fletcher L., Dennis B. R., Hudson H. S. et al., 2011, Space Sci Rev, 159, 19
\bibitem[2002]{Ghirlanda2002} Ghirlanda G., Celotti A., Ghisellini G., 2002, A\&A, 393, 49
\bibitem[2004]{Ghirlanda2004} Ghirlanda G., Ghisellini G. and Celotti A., 2004,  A\&A, 422, L55
\bibitem[2009]{Godet2009} Godet O., Sizun P. et al., 2009, Nucl. Instr. Meth. A, 603, 365.
\bibitem[1991]{Horack1991} Horack J. M., Fishman G. J., Meegan C. A. et al., 1991, AIP Conf. Proc., 265, 373
\bibitem[2006]{Kaneko2006} Kaneko Y., Preece B. D., Briggs M. S. et.al., 2006, ApJS, 166, 298
\bibitem[1992]{Kouveliotou1992} Kouveliotou C., Norris J. P., Wood K. S. et al., 1992, ApJ, 392, 179
\bibitem[2011]{Lin2011} Lin R. P., 2011, Space Sci. Rev., 159, 421
\bibitem[1993]{Mallozzi1993} Mallozzi R. S., Paciesas W. S., Meegan C. A. et al., 1993, AIP Conf. Proc., 280, 1122
\bibitem[2008]{Mandrou2008} Mandrou, P. Schanne S. et al., 2008, AIP Conf. Proc., 1065, 338
\bibitem[2004]{Mclean2004} Mclean K. M., Fenimore E.E., Palmer D. et al., A.2004, in AIP Conf. Proc., Vol.727, Gamma-Ray Burst: 30 Years of Discovery, ed. E.E.Fenimore and M.Galassi(Melville:AIP), p. 667
\bibitem[1993]{Meegan1993} Meegan C., Fishman G., Wilson R. and Paciesas W., 1993, AIP Conf. Proc., 280, 1117
\bibitem[2011]{Nava2011} Nava L., Ghirlanda G., Ghisellini G. and Celotti A., 2011, A\&A, 530, A21
\bibitem[2011]{Paul2011} Paul J., We J., Basa S., and Zhang S.-N., 2011, C. R. Phys., 12, 298
\bibitem[2008]{Schanne2008} Schanne S., Cordier B., G\"{o}tz D., 2008, 30th international cosmic ray conference, 3(OG part 2), 1147
\bibitem[2001]{Tavenner2001} Tavenner T.anya, Ed Fenimore, Mark Galassi et al.,  2003, AIP Conf. Proc., 662, 97
\bibitem[2012]{Zhao2012} Zhao D. H., CORDIER B., Sizun P. et al., 2012, Exp Astron, Vol. 34, Issue 3, 705
\end{thebibliography}
\end{document}